\documentclass[journal=jpclcd,manuscript=letter]{achemso}

\usepackage{chemformula} % Formula subscripts using \ch{}
\usepackage[T1]{fontenc} % Use modern font encodings
\usepackage{orcidlink}
\usepackage{graphicx}
\usepackage{caption}
\usepackage{subcaption}
\usepackage{romanbar}
\usepackage{booktabs}
\usepackage{hyperref}
\usepackage{subcaption}
\usepackage{tabularx}
\usepackage{siunitx}
\usepackage{multirow}
\usepackage{amssymb}
\usepackage{comment}

\usepackage{verbatim}
\newcommand{\Rom}[1]{\uppercase\expandafter{\romannumeral #1\relax}}

\author{Jannik Theile\orcidlink{0009-0003-3708-1342}}
\affiliation{Carl von Ossietzky Universit\"at Oldenburg, Institute of Physics, 26129 Oldenburg, Germany}
\email{jannik.theile@uni-oldenburg.de}

\author{Michele Guerrini\orcidlink{0000-0002-9652-8832}}
\affiliation{Friedrich-Schiller-Universit\"at Jena, Institut f\"ur Festk\"orpertheorie und -Optik, 07743 Jena, Germany}
\email{michele.guerrini@uni-jena.de}

\author{Caterina Cocchi\orcidlink{0000-0002-9243-9461}}
\affiliation{Friedrich-Schiller-Universit\"at Jena, Institut f\"ur Festk\"orpertheorie und -Optik, 07743 Jena, Germany}
\alsoaffiliation{Friedrich-Schiller-Universit\"at Jena, Abbe Center of Photonics, 07745 Jena, Germany}
\email{caterina.cocchi@uni-jena.de}

\title{Strong-Field-Driven Non-Linear Vibronic Coupling in an Asymmetric Donor-Acceptor Oligomer}


\begin{document}

\begin{tocentry}
\centering
\includegraphics[width=\textwidth]{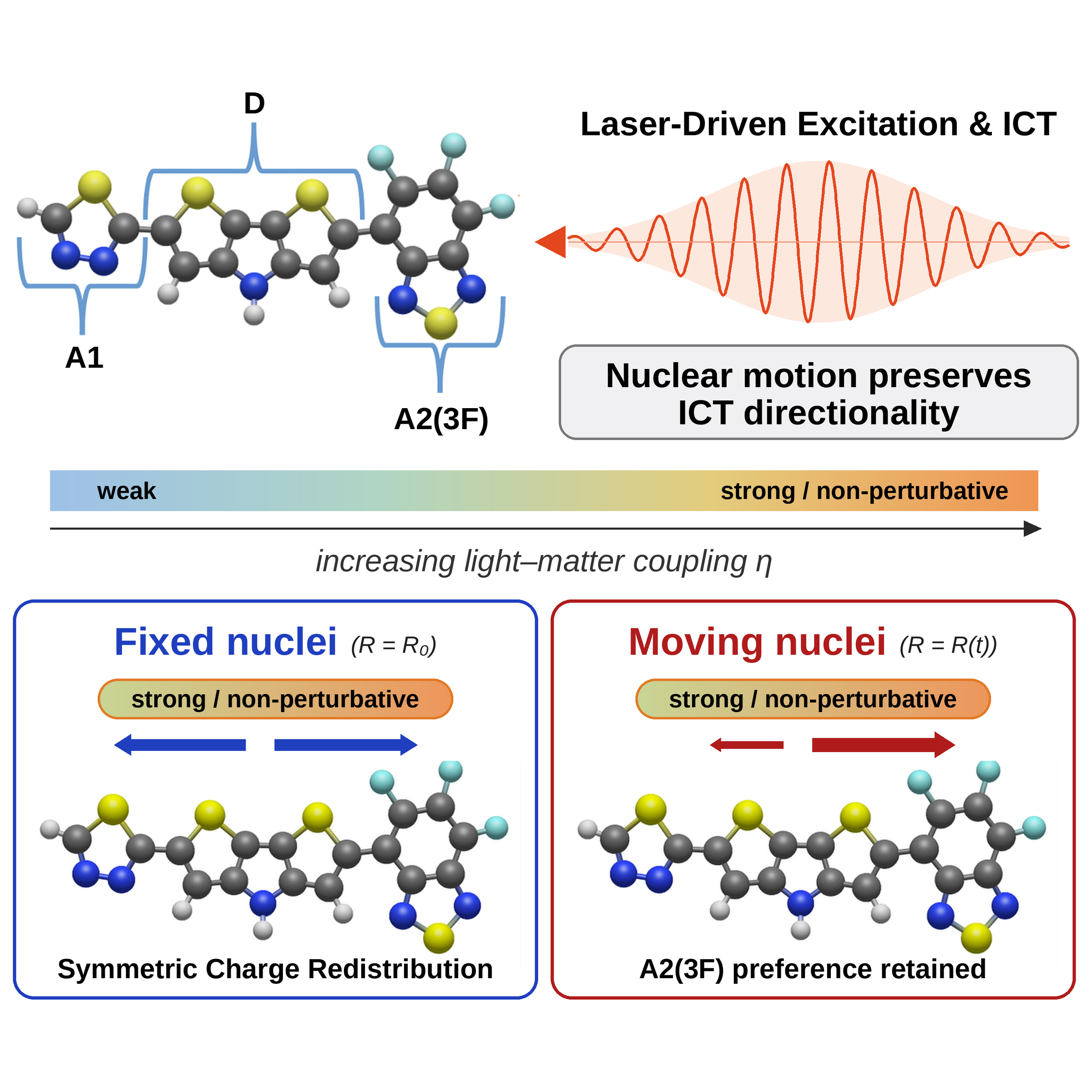}
\end{tocentry}

%%%%%%%%%%%%%%%%%%%%%%%%%%%%%%%%%%%%%%%%%%%%%%ƒ%%%%%%%%%%%%%%%%%%%%%%%
\begin{abstract}
Understanding how nuclear motion couples to intramolecular charge transfer beyond the perturbative regime of light--matter interaction remains an open problem. We tackle this problem by investigating the electron-nuclear response of an acceptor-donor-acceptor oligomer with structural asymmetry due to the partial fluorination of an acceptor unit. Adopting real-time time-dependent density functional theory combined with Ehrenfest dynamics, we excite the systems with a resonant pulse of varying intensity, spanning the weak-, intermediate-, and strong-coupling regimes. When nuclei are clamped, field-driven charge preferentially flows from the donor to the fluorinated acceptor, with selectivity reduced upon increasing intensity. With moving ions, the early-time spatial response remains qualitatively unchanged, despite a distinct, spatially asymmetric charge redistribution. In the strong-field regime, nuclear motion preserves the preference for field-driven charge accumulation on the fluorinated acceptor, although the process takes approximately twice the time as in the fixed-nuclei scenario. Mode-resolved kinetic-energy analysis reveals that, with increasing field strength, the laser-induced nuclear response shifts from selected bond stretchings to collective low-frequency modes. These results provide a benchmark for assessing the interplay between vibronic dynamics and strong-field coupling in donor/acceptor oligomers.
\end{abstract}

%\newpage

%%%%%%%%%%%%%%%%%%%%% Introduction %%%%%%%%%%%%%%%%%%%%% 

%\section{Introduction}
Photoinduced intramolecular charge transfer (ICT) in covalently bound donor-acceptor (D-A) compounds~\cite{pand+12mm,min+18oe} represents the key operational mechanism of organic photovoltaics, molecular electronics, and light-harvesting assemblies~\cite{Samanta2023,Goes2003,li+23am,Cheng2024,Rajamalli2016}. Existing studies have mostly focused on the weak light-matter coupling regime~\cite{desi+21natn,wint+25natc}, in which electronic transitions are well described by perturbative theories~\cite{nelson2018coherent,timmer2022charge,mukazhanova2023impact,winte2025vibronic}. Emerging strong-field protocols for probing molecular complexes in the femtosecond or even attosecond time scales~\cite{miller2016laser,nisoli2017attosecond,alexander2025attosecond} require, however, a deeper understanding of the physical mechanisms ruling the coupling between structural, electronic, and electromagnetic degrees of freedom.
Excitations driven by intense, resonant laser pulses open non-perturbative pathways, such as multi-photon transitions or Stark-shifted state mixing~\cite{Autler1955Stark,trallero2005coherent,sussman2006dynamic}. These effects drive a massive charge separation far beyond linear boundaries, fundamentally altering the electronic density along the covalent backbone of a single complex~\cite{Rozzi_2017, Tamura2007PhononDriven}. 

Moving away from the perturbative regime, in which ionic motion acts as a small perturbation modulating coherent electronic oscillations via harmonic normal modes, the rapid displacement of the charge density driven by intense radiation dramatically alters the potential energy surface on a sub-femtosecond timescale~\cite{yats-guer02pra,Ivanov2024SBCT}. As a consequence, the nuclei are subjected to sizable localized forces, inducing structural deformations substantially altering electronic propagation. The study of electron-nuclear dynamics in the non-perturbative regime has historically been restricted to small, gas-phase molecules~\cite{mark+07pra,Zhang_2020}, where explicit solutions of the time-dependent Schr\"odinger equation are computationally accessible~\cite{aebersold2019coupled,pandey2023strong}. Investigating realistic conjugated systems under strong field irradiation requires suitable theoretical approaches capturing structural changes alongside electronic quantum dynamics at affordable costs. 

Real-time time-dependent density functional theory (RT-TDDFT)~\cite{marq-gros04arpc,li2020real} in tandem with Ehrenfest molecular dynamics~\cite{andrade2009modified, Lie_Ehrenfest, Rozzi_2017} provides an efficient first-principles framework to accomplish this task. The main advantage of this approach is allocating the quantum-mechanical treatment entirely on the electron dynamics while describing the nuclear motion via semiclassical trajectories bypassing the exponential scaling of the Schr\"odinger equation~\cite{Rozzi_2017}. This non-perturbative scheme additionally allows dynamical tuning of the field intensity to explore various regimes of excitation~\cite{castro2004excited,determination_linear_spectrum,krum+24apr} without any \textit{a priori} assumptions regarding the order of the relevant process or expansion truncations~\cite{vang-snij98jcp,park+18jctc}. Yet, how strong light-matter couplings and electron-nuclear interactions affect charge dynamics in conjugated systems remains an open question.

\begin{figure*}
    \centering
    \includegraphics[width=\linewidth]{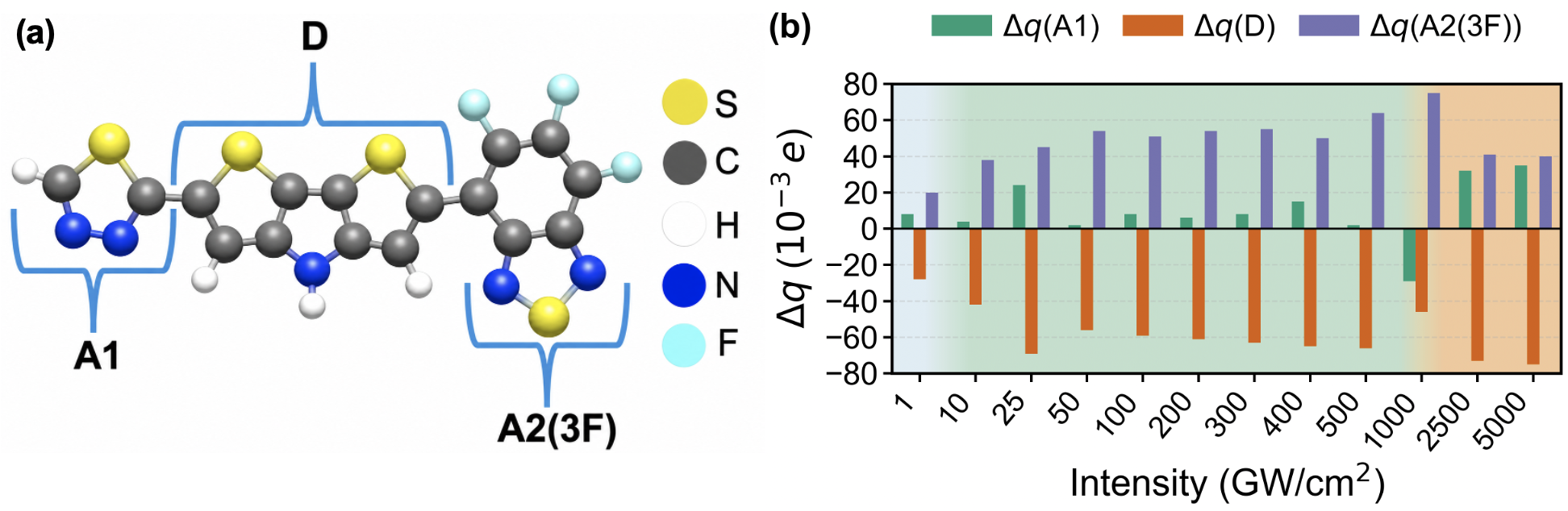}
    \caption{a) Ball-and-stick representation of the covalently bound acceptor-donor-acceptor oligomer consisting of a central dithieno[3,2-b:2',3'-d]pyrrole (DTP) donor core coupled to two distinct terminal units: a 1,3,4-thiadiazole (TD, labeled A1) and a triply fluorinated benzo[c][1,2,5]thiadiazole [BTD, labeled A2(3F)]. (b) Intensity-dependent time-averaged charge redistribution $\Delta q$ under frozen-ion (fixed-nuclei) conditions. Shaded backgrounds identify the weak/linear (light blue), intermediate (light green), and strong/non-perturbative (orange) excitation regimes, with smooth transitions around $\eta_\mathrm{nl}\sim10^{-2}$ and $\eta_\mathrm{np}\approx0.16$--$0.17$.}
    \label{fig:overview}
\end{figure*}

In this Letter, we unravel the interplay between strong-field electron dynamics and real-time nuclear motion in an asymmetric, covalently bound acceptor-donor-acceptor (A-D-A) oligomer consisting of a central dithieno[3,2-b:2',3'-d]pyrrole (DTP) donor coupled to two distinct terminal units: a 1,3,4-thiadiazole (TD, labeled A1) and a triply fluorinated benzo[c][1,2,5]thiadiazole [BTD, labeled A2(3F)] (Figure~\ref{fig:overview}a). As established in previous work~\cite{Theile_ICT}, targeted fluorination of a single acceptor segment maximizes electronic unbalance across the backbone, yielding spatially separated frontier orbitals and a steep potential gradient along the conjugation network. Using RT-TDDFT in conjunction with Ehrenfest dynamics, we monitor fragment-resolved charge redistribution across weak, intermediate, and strong non-perturbative light-matter coupling regimes quantified via a general, dimensionless parameter. By comparing fixed- and moving-nuclei trajectories, we demonstrate that explicit nuclear motion is essential for preserving the directional preference for charge accumulation on the fluorinated acceptor under strong-field excitation. Furthermore, mode-resolved kinetic-energy analysis reveals that this vibronic stabilization is governed by an intensity-driven shift from selected bond stretchings to collective low-frequency modes. Together, these results provide a robust benchmark for assessing the intricate interplay between vibronic dynamics and strong-field interactions in donor-acceptor oligomers.

%%%%%%%%%%%%%%%%%%%%%%%%%%%%%%%%%%%%%%%%%%%%%%%%%%%%%%
%%%%%%%%%%%%%%%%%%%%%%%%%%%%%%%%%%%%%%%%%%%%%%%%%%%%%%

To establish a generalizable framework for the light-driven dynamics across different excitation regimes, we introduce the dimensionless light-matter coupling parameter~\cite{Driouech2026, Autler1955Stark,Scully1997,CohenTannoudji1998}
\begin{equation}
\eta = \frac{|{\bf E}_0\cdot{\boldsymbol\mu}_{0I}|}{\hbar\omega_\mathrm{L}},
\label{eq:coupling_param}
\end{equation}
which is directly proportional to the absolute product of the peak electric-field amplitude of the Gaussian pulse (${\bf E}_0$) and the transition dipole moment of the resonantly pumped excitation in the linear-response spectrum of the molecule (${\boldsymbol\mu}_{0I}$), and to the inverse of the laser pump driving frequency $\omega_\mathrm{L}$. 
To classify the excitation regimes explored in this work, we quantify the energy retained after the pulse in photon-number units, $N_\mathrm{ph}=\Delta E/\hbar\omega_\mathrm{L}$, which allows us to define two operational scenarios. The deviation of $N_\mathrm{ph}$ by more than $10\%$ from the perturbative one-photon linear scaling behavior, $N_\mathrm{ph}\propto\eta^2$, marks the \textit{nonlinear threshold} $\eta_\mathrm{nl}$, while the \textit{non-perturbative} threshold $\eta_\mathrm{np}$ defines the limit in which the retained energy reaches that of at least two resonant photons, $N_\mathrm{ph}\geq2$. 

In our study, the values $\eta_\mathrm{nl}\sim10^{-2}$ ($I\approx3~\mathrm{GW\,cm^{-2}}$) and $\eta_\mathrm{np}=0.16$--$0.17$ ($I\approx1~\mathrm{TW\,cm^{-2}}$) mark gradual crossovers visualized by color transitions in Figure~\ref{fig:overview}b. Below $\eta_\mathrm{nl}$, the system operated in the weak-coupling regime, where energy uptake scales linearly with the field strength up to $\eta=0.01$. Intermediate driving ($0.01<\eta<0.16$) leads to a plateau in the energy uptake, which subsequently steepens at the onset of the strong-coupling regime around $\eta_\mathrm{np}=0.16$-$0.17$ (Figure S2). For $\eta=0.26$--$0.37\gtrsim\eta_\mathrm{np}$, the retained energy exceeds the two-photon equivalents, defining an increasingly non-perturbative response. A complete mapping of the intensities to $\eta$ is provided in Tables~S1 and S2. 

To establish a baseline for light-driven charge transfer, we isolate the purely electronic response of the A1--D--A2(3F) oligomer (Figure~\ref{fig:overview}a) by clamping the nuclei to their ground-state equilibrium configuration. The resulting time-averaged fragment charge variation,
\begin{equation}
\langle \Delta q_k \rangle_t 
= \frac{1}{T} \int_{0}^{T} \Delta q_k(t)\,dt,
\end{equation}
quantifies how the resonant field redistributes electronic density across the molecular backbone. Here, $T$ denotes the total propagation time, and $\Delta q_k(t)=\sum_{n\in k}\Delta q_n(t)$ represents the cumulative charge variation on fragment $k$, computed by summing atomic charge differences $\Delta q_n(t)=q_n(t)-q_n(0)$ over all constituent atoms $n$ via the smooth real-space partitioning scheme detailed in the Methodology section.

Under weak-field conditions ($I=1~\mathrm{GW\,cm^{-2}}$, $\eta \approx 0.005$), the fixed-nuclei simulations reveal the expected trend (Figure~\ref{fig:overview}b, light blue shaded area): the central donor is depleted ($\langle \Delta q_{\mathrm{D}} \rangle_t < 0$), while electronic charge preferentially accumulates on the fluorinated acceptor $\mathrm{A2(3F)}$ rather than on $\mathrm{A1}$ ($\langle \Delta q_{\mathrm{A2(3F)}} \rangle_t \approx +0.020~e$ vs.\ $\langle \Delta q_{\mathrm{A1}} \rangle_t \approx +0.008~e$). As the field intensity increases into the intermediate regime (green shaded area in Figure~\ref{fig:overview}b), this directional asymmetry becomes pronounced. Charge accumulation on $\mathrm{A2(3F)}$ scales up significantly, peaking at $\approx +0.075~e$ near the upper boundary ($\eta \approx 0.163$), whereas $\mathrm{A1}$ remains largely unaffected or exhibits minor fluctuations. This demonstrates that the static electronic potential gradient inherently funnels field-driven charge toward the fluorinated terminal unit. However, a qualitative shift occurs in the strong-field regime ($I\ge 2.5~\mathrm{TW\,cm^{-2}}$): while donor depletion remains sizeable ($\approx-0.07~e$), charge accumulation on $\mathrm{A2(3F)}$ drops to $+0.04~e$ while $\mathrm{A1}$ simultaneously gains $+0.03$ to $+0.04~e$. Thus, under these extreme clamped-nuclei conditions, fragment selectivity breaks down, and field-driven charge redistributes almost evenly between both terminal units.

\begin{figure*}
\centering
\includegraphics[width=\textwidth]{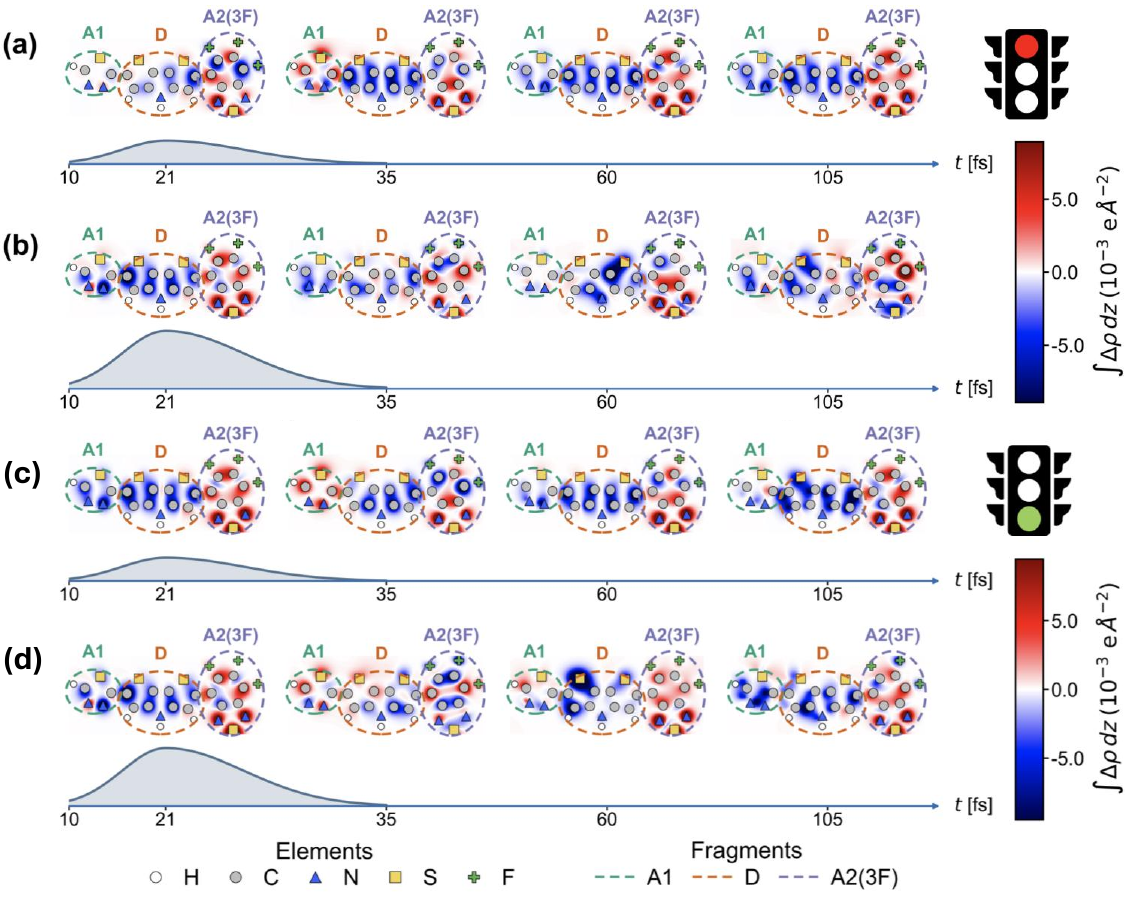}
\caption{Induced charge density maps of the A1--D--A2(3F) oligomer
projected onto the molecular plane at selected propagation times
during and after laser pulses of two different intensities
with (a)-(b) fixed and (c)-(d) moving nuclei. In both scenarios, two excitation regimes are explored: (a),(c) weak-coupling 
($\eta \approx 0.005$, $I=1~\mathrm{GW\,cm^{-2}}$) and (b),(d) upper-intermediate coupling ($\eta \approx 0.115$, $I=500~\mathrm{GW\,cm^{-2}}$).
Snapshots are evaluated at the pulse peak ($t \approx 21~\mathrm{fs}$), when the excitation ends ($t \approx 35~\mathrm{fs}$), and at later times after irradiation ($t=60$ and $105~\mathrm{fs}$). The projected density $\int\Delta\rho\,dz$ is obtained by integrating along the axis perpendicular to the molecular plane. Red and blue regions indicate electron accumulation
and depletion, respectively.}
\label{fig:2D}
\end{figure*}

To map the spatial distribution of the field-driven electron dynamics under clamped-nuclei conditions (Figure~\ref{fig:2D}a,b), we examine snapshots of the induced density $\Delta\rho$ (Eq.~\ref{eq:induced_density}) projected onto the molecular plane across key temporal windows: the pulse peak ($t = 21\,\mathrm{fs}$), the end of irradiation ($t = 35\,\mathrm{fs}$), and post-pulse evolution ($t = 60$ and $105\,\mathrm{fs}$; Figure~\ref{fig:2D} and Tables~S3--S4). Under clamped-nuclei conditions (Figure~\ref{fig:2D}a,b), the weak-coupling regime ($I=1~\mathrm{GW\,cm^{-2}}$) reveals localized charge accumulation (red) on the sulfur and nitrogen centers of $\mathrm{A2(3F)}$, while the donor core undergoes depletion (blue). The fluorine substituents (green crosses) remain largely unpopulated, supporting an inductive role: edge fluorination lowers the local $\pi$-electron density, turning the sulfur- (yellow squares) and nitrogen-rich (blue triangles) regions into a preferred electronic sink. After the pulse, this accumulation decays as charge flows back toward $\mathrm{D}$ and $\mathrm{A1}$. In the upper-intermediate regime ($I=500~\mathrm{GW\,cm^{-2}}$), the initial spatial pattern is identical, but the response exhibits a sublinear, saturating scaling: despite a 500-fold higher intensity, the induced density increases only moderately. Rather than relaxing, the post-pulse charge distribution undergoes coherent oscillations, alternating between accumulation and depletion on $\mathrm{A2(3F)}$ at $35$, $60$, and $105\,\mathrm{fs}$. This represents a characteristic behavior of unperturbed electronic wavepacket dynamics in the absence of nuclear relaxation~\cite{jaco+23jpca}.

Allowing the nuclei to move within the Ehrenfest framework fundamentally alters this post-pulse evolution (Figure~\ref{fig:2D}c,d). In the weak-coupling regime (Figure~\ref{fig:2D}c), nuclear motion exerts minimal influence up to $t=35\,\mathrm{fs}$, modifying only late-time charge distributions. However, in the upper-intermediate regime ($I=500~\mathrm{GW\,cm^{-2}}$, Figure~\ref{fig:2D}d), electron--nuclear coupling disrupts the coherent reversibility seen in the fixed-nuclei case. While early-time accumulation ($t=21\,\mathrm{fs}$) mimics the clamped-nuclei reference, the migration toward $\mathrm{A2(3F)}$ remains pronounced at $t=35\,\mathrm{fs}$, where it weakens for clamped nuclei. Subsequently, nuclear motion channels density away from the fluorinated periphery of $\mathrm{A2(3F)}$ toward its sulfur and nitrogen centers. By $t=105\,\mathrm{fs}$, this feedback yields a spatially asymmetric redistribution pattern rather than the uniform fragment-wide oscillations of the fixed-nuclei scenario. This decoupling confirms that structural degrees of freedom actively reshape post-pulse charge dynamics, preventing simple electronic back-flow and stabilizing the directional charge distribution.

\begin{figure*}
    \centering
    \includegraphics[width=\textwidth]{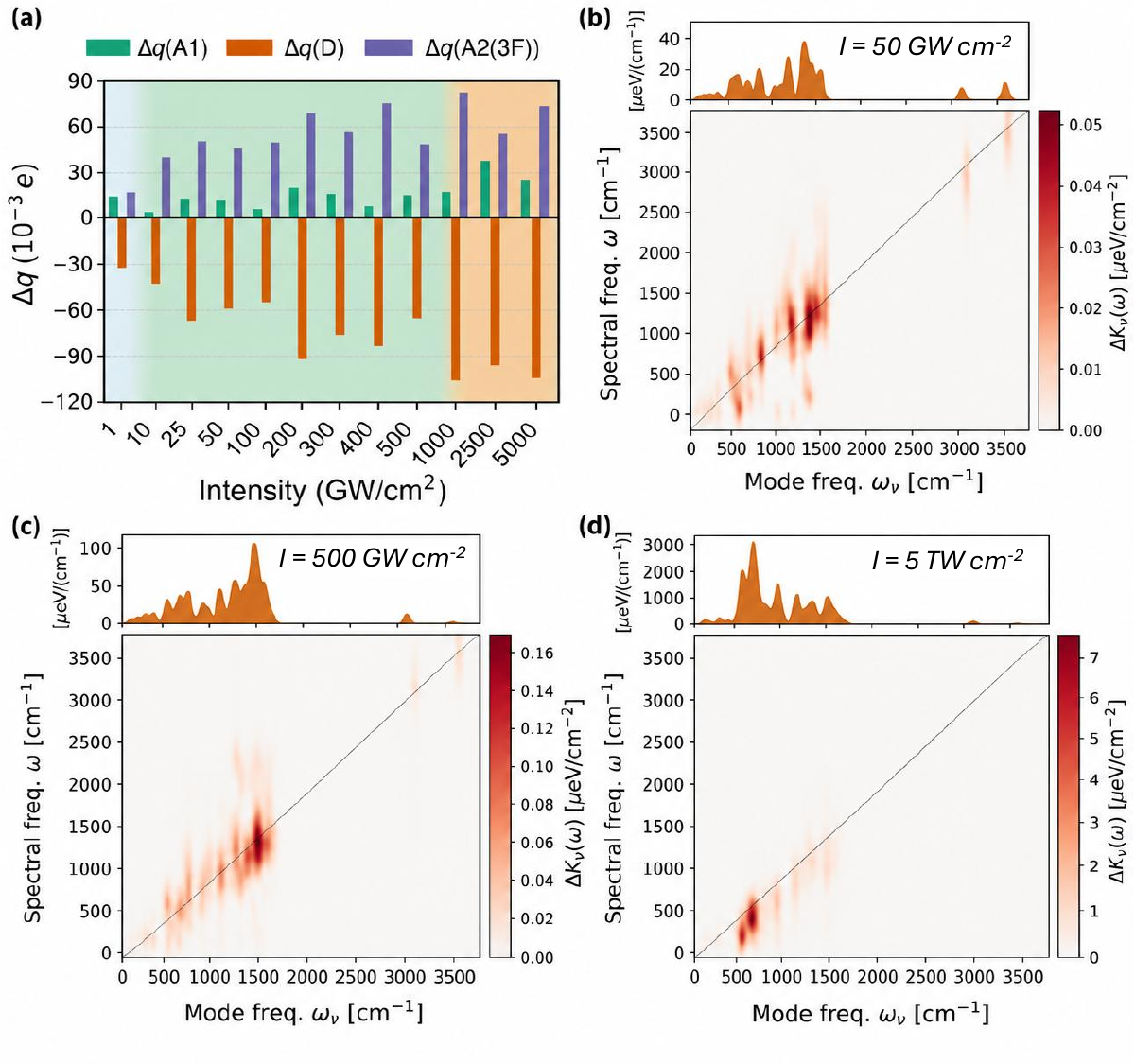} 
    \caption{(a) Time-averaged charge redistribution $\Delta q$ for moving nuclei as a function of intensity. Shaded regions mark the weak/linear (light blue), intermediate (light green), and strong/non-perturbative (orange) regimes, with transitions around $\eta_\mathrm{nl}\sim10^{-2}$ and $\eta_\mathrm{np}\approx0.16$--$0.17$. (b)--(d) Laser-induced vibrational kinetic-energy difference maps at (b) lower-intermediate $I=50~\mathrm{GW\,cm^{-2}}$ ($\eta\approx0.036$), (c) upper-intermediate $I=500~\mathrm{GW\,cm^{-2}}$ ($\eta\approx0.115$), and (d) non-perturbative $I=5~\mathrm{TW\,cm^{-2}}$ ($\eta\approx0.364$) excitation regimes as a function of the harmonic normal-mode frequency $\omega_{\nu}$ and the spectral frequency $\omega$ associated to the spectral distribution of the kinetic energy of each normal mode. Positive (negative) values indicate an increase (decrease) in vibrational kinetic energy relative to the field-free dynamics; the dashed line marks $\omega=\omega_{\nu}$. Upper panels: frequency-integrated change per normal mode.}
\label{fig:vibronic_analysis}
\end{figure*}

Since the fixed- and moving-nuclei simulations share identical initial electronic states and field conditions, the emerging divergence in their post-pulse dynamics highlights the crucial role of vibronic coupling. Nuclear motion disrupts the coherent, purely electronic charge oscillations of the clamped-nuclei reference, producing a partial spatial redistribution at late propagation times. While transient snapshots (Figure~\ref{fig:2D}) reveal a late-time decay on $\mathrm{A2(3F)}$, time-averaged fragment charges (Figure~\ref{fig:vibronic_analysis}a) confirm that early-time transfer dominates, preserving directional donor-to-fluorinated-acceptor ICT across all excitation regimes. Specifically, while the fixed-nuclei response breaks down into symmetric terminal sharing at extreme field conditions (Figure~\ref{fig:overview}b), nuclear motion maintains a pronounced preference for $\mathrm{A2(3F)}$. Across the non-perturbative regime, the central donor undergoes persistent depletion (scoring up to $-0.103\,e$ at $I=1~\mathrm{TW\,cm^{-2}}$), while $\mathrm{A2(3F)}$ accumulates significant electron density, peaking at $+0.084\,e$ at $I=1~\mathrm{TW\,cm^{-2}}$ and recovering to $+0.074\,e$ at $I=5~\mathrm{TW\,cm^{-2}}$ following a transient competitive dip at $I=2.5~\mathrm{TW\,cm^{-2}}$ where $\mathrm{A1}$ temporarily captures a larger share. At $5~\mathrm{TW\,cm^{-2}}$, nuclear motion nearly doubles the time-averaged accumulation on $\mathrm{A2(3F)}$ relative to the clamped-nuclei scenario ($+0.074\,e$ vs.\ $\approx+0.04\,e$, a factor of $\approx1.9$), compared to a modest effect at $1~\mathrm{TW\,cm^{-2}}$ ($+0.084\,e$ vs.\ $+0.075\,e$). This late-time divergence indicates strong dynamical feedback, in which field-driven electronic density displacements exert localized forces that drive nuclear motion. 

To uncover the specific structural coordinates mediating this feedback and stabilizing the directional charge transfer, we perform a mode-resolved vibrational kinetic-energy analysis~\cite{Herperger2021}, monitoring differential maps visualizing $\Delta K_{\nu}(\omega) = K_{\nu}^{\mathrm{laser}}(\omega) - K_{\nu}^{\mathrm{free}}(\omega)$, where $K_{\nu}^{\mathrm{laser}}$ and $K_{\nu}^{\mathrm{free}}$ represent the kinetic energy spectrum of the laser-excited system and the corresponding field-free baseline, respectively (Figure~\ref{fig:vibronic_analysis}b--d and Figure~S5). Deviations from the diagonal, where harmonic modes reside, refer to excitation-induced red- or blue-shifted components.
Across all intensities, energy uptake concentrates below $\tilde{\nu}=1600~\mathrm{cm^{-1}}$ in conjugated backbone deformations, with C--H and N--H stretching modes above $3000~\mathrm{cm^{-1}}$ remaining negligible (Figure~S6). Crucially, increasing laser intensities drive a non-uniform spectral reorganization rather than a simple thermal redshift. In the intermediate-coupling regime (Figure~\ref{fig:vibronic_analysis}b,c), the response is dominated by high-frequency backbone stretches ($\approx 1400\text{--}1600~\mathrm{cm^{-1}}$\, coupled C--C, C--N and C--F stretches, Figure~S6). At the onset of the non-perturbative regime ($I=1~\mathrm{TW\,cm^{-2}}$), spectral weight shifts toward the $800\text{--}950$ and $500\text{--}700~\mathrm{cm^{-1}}$ manifolds involving sulfur-containing heterocycles, culminating at $5~\mathrm{TW\,cm^{-2}}$ in a response dominated by the $500\text{--}700~\mathrm{cm^{-1}}$ C--S stretching modes of the thiadiazole acceptor ring (Figure~\ref{fig:vibronic_analysis}d and Figure~S6a). Concurrently, the frequency-integrated response $\Delta\langle K_{\nu}\rangle$ transitions from sublinear scaling (increasing by only a factor of $\sim 2.5$ between $50$ and $500~\mathrm{GW\,cm^{-2}}$) to superlinear growth (jumping by a factor of $\sim 30$ at $I=5~\mathrm{TW\,cm^{-2}}$) upon a further tenfold increase.

The pronounced spectral shifts from high-frequency backbone stretching to collective low-frequency motion coincide with the intensity window in which directional charge separation is enhanced relative to the clamped-nuclei reference. This evidence demonstrates that vibronic stabilization is not driven by indiscriminate nuclear motion, but by an intensity-dependent reorganization into collective sulfur/ring deformations at frequencies ranging between 500 and 700~$\mathrm{cm^{-1}}$. At extreme field strengths, charge separation saturates despite maximum nuclear redistribution, underscoring the delicate, non-linear interplay between electronic driving and structural trapping.

In summary, we have unravelled the real-time electron-nuclear dynamics of an asymmetric acceptor--donor--acceptor oligomer across the weak, intermediate, and strong-field regimes. Under fixed-nuclei conditions, laser-driven charge transfer follows the intrinsic electrostatic gradient toward the triply fluorinated acceptor $\mathrm{A2(3F)}$, primarily accumulating at its sulfur and nitrogen centers while leaving the fluorine substituents largely unpopulated. However, this directionality degrades at extreme field strengths, where charge accumulation redistributes nearly symmetrically between both terminal units. Activating nuclear motion within the Ehrenfest framework fundamentally alters this picture. In the non-perturbative regime, electron--nuclear coupling suppresses post-pulse electronic oscillations, replacing them with a spatially asymmetric redistribution that preserves the strong directional preference for $\mathrm{A2(3F)}$, nearly doubling the time-averaged charge retention relative to the clamped-nuclei reference. Mode-resolved analysis reveals that this vibronic stabilization is driven by an intensity-dependent spectral reorganization: as the field strength increases, the laser-induced nuclear response shifts from high-frequency backbone stretching modes into collective low-frequency deformations involving sulfur-containing heterocycles. At extreme field intensities, charge separation saturates despite maximum nuclear activity, underscoring a nonlinear interplay between electronic driving and structural trapping. By establishing how low-frequency structural reorganization preserves directional charge separation under intense radiation, these findings provide a foundational benchmark for designing nonlinear organic optoelectronic materials.

%%%%%%%%%%%%%%%%%%%%%%APPENDIX%%%%%%%%%%%%%%%%%%%
\section{Methodology}
\subsection{Theoretical Background}
The results presented in this study are obtained in the framework of RT-TDDFT coupled with Ehrenfest molecular dynamics. The time evolution of the electronic degrees of freedom is governed by the time-dependent Kohn-Sham equation~\cite{Runge_Gross}:
\begin{equation}
    i\hbar \frac{\partial}{\partial t}\psi_n(\mathbf{r},t) = \hat{H}_{\mathrm{KS}}[\rho](\mathbf{r},t)\psi_n(\mathbf{r},t),
\end{equation}
where the effective single-particle Hamiltonian,
\begin{equation}
    \hat{H}_{\mathrm{KS}}[\rho](\mathbf{r},t) = -\frac{\hbar^2}{2m_e}\nabla^2 + V_{e-n}(\mathbf{r},\mathbf{R}(t)) + V_H[\rho](\mathbf{r},t) + V_{xc}[\rho](\mathbf{r},t) + V_{\mathrm{ext}}(\mathbf{r},t),
\end{equation}
accounts for the kinetic energy, electron--nuclear interactions, the classical Hartree potential, the exchange-correlation potential, and the external driving potential, respectively. The classical nuclei evolve simultaneously on a mean-field potential energy surface driven by the Ehrenfest force, obtained from the expectation value of the gradient of the electronic Hamiltonian with respect to the nuclear coordinates $\mathbf{R}(t)$.

The interaction of the molecular system with the external laser field is treated in the length gauge within the dipole approximation:
\begin{equation}
    V_{\mathrm{ext}}(\mathbf{r},t) = -e\,\mathbf{r}\cdot\mathbf{E}(t),
\end{equation}
where the driving electric field is modeled as a monochromatic carrier modulated by a Gaussian envelope:
\begin{equation}
    \mathbf{E}(t) = \hat{\mathbf{n}} E_0 \exp\left[-\frac{(t-t_\mu)^2}{2t_\sigma^2}\right] \cos(\omega_\mathrm{L} t).
\end{equation}
 Here, $\hat{\mathbf{n}}$ denotes the polarization direction, $E_0$ the peak field amplitude, $t_\mu=21$~fs the pulse center, $t_\sigma=7$~fs the pulse width, and $\omega_\mathrm{L}$ the carrier frequency.

To quantify the dynamic, directional charge redistribution without grid-induced artifacts, we employ a smooth real-space partitioning scheme using distance-based soft weights. The molecular structure is divided into three functional fragments $k$: $\mathrm{A1}$, $\mathrm{D}$, and $\mathrm{A2(3F)}$. Each grid point $\mathbf{r}$ is assigned to a fragment $k$ via a Gaussian kernel function:
\begin{equation}
    w_k(\mathbf{r}) =\frac{\exp\!\left[-d_k^2(\mathbf{r})/
(2\sigma_\mathrm{w}^2)\right]}{\sum_j\exp\!\left[-d_j^2(\mathbf{r})/(2\sigma_\mathrm{w}^2)\right]},
\end{equation}
where $d_k(\mathbf{r}) = \min_{\alpha \in k} |\mathbf{r} - \mathbf{R}_\alpha|$ denotes the Euclidean distance to the nearest atom $\alpha$ belonging to fragment $k$, and $\sigma_\mathrm{w}$ controls the spatial smoothness of the boundaries. By construction, the weights satisfy the partition of unity $\sum_k w_k(\mathbf{r}) = 1$ at every point in space. The time-dependent fragment populations $q_k(t)$ are then evaluated via a discrete summation over the grid voxels of volume $\Delta V$:
\begin{equation}
    q_k(t) \approx \sum_i w_k(\mathbf{r}_i)\,\rho(\mathbf{r}_i,t)\,\Delta V,
\end{equation}
from which $\Delta q_k(t)=q_k(t)-q_k(0)$ follows. This provides a numerically stable representation of continuous charge flow across the molecular subunits.

The laser-induced redistribution of electronic charge is visualized by the
\emph{induced density}, i.e.\ the change of the total electronic density with respect to the ground state,
\begin{equation}
    \Delta\rho(\mathbf{r},t) = \rho(\mathbf{r},t) - \rho_{\mathrm{GS}}(\mathbf{r}),
    \qquad
    \rho(\mathbf{r},t) = \sum_{n}^{\mathrm{occ}} \left|\psi_n(\mathbf{r},t)\right|^2.
    \label{eq:induced_density}
\end{equation}

\subsection{Computational Details}
All simulations were performed using the real-space grid code \texttt{Octopus} (version 13.0)~\cite{Octopus2}. The time-dependent electron density was represented on a grid defined by the union of atom-centered spheres with a radius of 8~\AA{} and a grid spacing of 0.12~\AA{}, ensuring rigorous convergence of both total energies and transient response properties. Ground-state geometry optimizations were carried out using the FIRE algorithm~\cite{FIRE_optimization} until the residual forces on all atoms fell below $10^{-2}$~eV/\AA{}, using a self-consistent field convergence criterion of $10^{-8}$ for the relative density. Core electrons were modeled using Hartwigsen--Goedecker--Hutter pseudopotentials~\cite{Hartwigsen_Goedecker_Hutter}. 

The exchange-correlation potential was described within the local-density approximation (LDA) with the Perdew-Zunger parametrization~\cite{Perdew_Zunger}. While local functionals omit long-range asymptotic corrections and can underestimate static charge-transfer excitation energies, they offer a computationally tractable and numerically robust framework for propagating electronic wavefunctions on real-space grids over extended timescales. In the context of Ehrenfest dynamics, where non-local exchange operators become prohibitively expensive due to the dense sub-femtosecond sampling required for the electronic wavefunctions, the adiabatic LDA framework serves as a reliable baseline to capture the qualitative mechanisms of coherent electronic redistribution and the initial stages of dynamically driven charge transfer. We emphasize that the combination of adiabatic LDA and mean-field (Ehrenfest) nuclear dynamics neglects decoherence and nuclear branching; the results should therefore be understood as a qualitative description of the coupled electron--nuclear response, and hybrid functionals and beyond-mean-field nuclear dynamics remain to be tested.

For the real-time dynamics, the central driving frequency was set to $\hbar\omega_\mathrm{L}=1.80~\mathrm{eV}$, corresponding to the primary low-energy absorption peak obtained from the linear optical response via an isotropic $\delta$-kick perturbation \cite{Yabana_Bertsch_1996} and verified through Casida's formalism \cite{Casida1995}. Given that this transition is predominantly polarized along the long molecular axis ($x$-axis), the incoming laser fields were oriented parallel to this direction. Laser intensities spanning from $1$~GW/cm$^2$ to $5$~TW/cm$^2$ were employed. The total propagation time $T$ was 105~fs, executed with an electronic time step of 1~as. Time evolution was tracked using the approximated enforced time-reversal symmetry (AETRS) propagator, while the action of the exponential time-evolution operator was evaluated using a Lanczos scheme \cite{lanczos}. Parallel control simulations were performed with a clamped lattice to explicitly isolate the influence of vibronic coupling.

 Vibrational normal modes were calculated for the optimized ground-state geometry using the Sternheimer linear-response approach implemented in \texttt{Octopus}~\cite{Octopus1}. The corresponding atomic displacement vectors were inspected to assign the dominant vibrational character of representative modes, including stretching contributions and collective in-plane and out-of-plane nuclear motions.
%%%%%%%%%%%%%%%%%%%%% End matter %%%%%%%%%%%%%%%%%%%%% 

%%%%%%%%%%%%%%%%%%%%% Acknowledgement %%%%%%%%%%%%%%%%%%%%% 
\begin{acknowledgement}
This work was financially supported by the Lower Saxony Ministry of Science and Culture (project ElLiKo) and by the German Research Foundation (DFG), project numbers 398816777 (CRC 1375 ``NOA'', subproject A8) and 579461218 (DYNOPEN). Computational resources are provided by the high-performance computing cluster ROSA at the University of Oldenburg, funded by the German Research Foundation (Project No. INST 184/225-1 FUGG) and by the Lower Saxony Ministry of Science and Culture.
\end{acknowledgement}

%%%%%%%%%%%%%%%%%%%%% SI info %%%%%%%%%%%%%%%%%%%%%
\begin{suppinfo}
The Supporting Information is available free of charge:
\begin{itemize}
    \item Linear-response absorption spectra, excitation-character analysis, transition-density analysis, and transition dipole moments obtained from RT-TDDFT and the Casida formalism;
    \item Definition and numerical evaluation of the dimensionless light--matter coupling parameter $\eta$, including energy-based criteria for nonlinear and non-perturbative thresholds and the corresponding classification of light--matter interaction regimes;
    \item Time-dependent fragment-resolved charge redistribution $\Delta q(t)$ for fixed and moving nuclei;
    \item Definition of the induced density and its decomposition into density-difference and transition-density contributions, and fragment-resolved induced charge redistribution ($\Delta q_k^{\mathrm{V}}$) at selected propagation times and laser intensities for fixed and moving nuclei;
    \item Computational details of the vibrational mode analysis, laser-induced vibrational kinetic-energy difference maps, and representative normal-mode displacement patterns.
\end{itemize}
\end{suppinfo}

\section*{Data Availability Statement}
The datasets underlying the main figures and conclusions presented in this work are freely available in the Zenodo repository at \href{https://doi.org/10.5281/zenodo.21885972}{DOI: 10.5281/zenodo.21885972}.  Due to their substantial file size, the complete trajectory histogram datasets are available from the corresponding author upon reasonable request. 

%%%%%%%%%%%%%%%%%%%%% competing interest declartion %%%%%%%%%%%%%%%%%%%%% 
\section*{Notes}
The authors declare no competing financial interest.

%%%%%%%%%%%%%%%%%%%%% biblography %%%%%%%%%%%%%%%%%%%%% 
%\bibliography{bib}
\providecommand{\latin}[1]{#1}
\makeatletter
\providecommand{\doi}
  {\begingroup\let\do\@makeother\dospecials
  \catcode`\{=1 \catcode`\}=2 \doi@aux}
\providecommand{\doi@aux}[1]{\endgroup\texttt{#1}}
\makeatother
\providecommand*\mcitethebibliography{\thebibliography}
\csname @ifundefined\endcsname{endmcitethebibliography}  {\let\endmcitethebibliography\endthebibliography}{}

\end{document}